\documentclass[fleqn,twoside]{article}
\usepackage{espcrc2}

\usepackage{graphicx}
\usepackage[figuresright]{rotating}

\def\NIMA#1#2#3{{\rm Nucl.~Instr.~and~Meth.} {\bf#1} (19#2) #3}
\def\PLB#1#2#3{{\it Phys.~Lett.} {\bf{B#1}} (19#2) #3}
\def\PRD#1#2#3{{\it Phys.~Rev.} {\bf{D#1}} (19#2) #3}
\def\PR#1#2#3{{\it Phys.~Rep.} {\bf{#1}} (#2) #3}
\def\NPB#1#2#3{{\rm Nucl.~Phys.} {\bf{B#1}} (19#2) #3}

\newcommand{\tbeta} {$\tan \beta$}
\newcommand{\HTT}{\tau\nu\tau\nu}
\newcommand{\HCSTN}{cs\tau\nu}
\newcommand{\HCSCS}{cscs}
\newcommand{\HWAWA}{W^*\!AW^*\!A}
\newcommand{\wa}{W^*\!A}
\newcommand{\HWATN}{W^*A\tau\nu}
\newcommand{\bb}{b\bar{b}}
\newcommand{\sLep}{\tilde{l}}
\newcommand{\sGra}{\tilde{G}}
\newcommand{\Chiz}{\tilde{\chi}^{0}}
\newcommand{\Chip}{\tilde{\chi}^{+}}
\newcommand{\Chim}{\tilde{\chi}^{-}}

\title{The use of the $\tau$ in new particle searches at DELPHI}

\author{G. G\'omez-Ceballos\address[MCSD]{Instituto de F\'{\i}sica de Cantabria (CSIC-UC), Avda.
     los Castros s/n, ES-39005 Santander, Spain}%
        \thanks{Now in the Massachusetts Institute of Technology}
                   }
       
\begin{document}

\begin{abstract}
Several new particle searches have been performed in the DELPHI experiment 
involving $\tau$ leptons in the resulting final state. The topology and 
special characteristics of the $\tau$ leptons have been used to discriminate 
the signal from the Standard Model background. Limits on new particles have 
been set, playing an important role the channels with $\tau$ leptons.

\vspace{1pc}
\end{abstract}

\maketitle

\section{Introduction}

$\tau$ leptons are not only a precision measurement tool, but also a sensitive probe for Physics beyond the
Standard Model. In this note we review the importance of their contribution and the results obtained at the
DELPHI experiment at LEP. Several extensions to the Standard Model predict the existence of new 
particles involving $\tau$ leptons in the resulting final state.

Data collected at LEP2 at centre-of-mass energies from 130 GeV 
to 209 GeV were used in the different analyses, with a total integrated 
luminosity of about 650 pb$^{-1}$.

\section{The DELPHI detector}

DELPHI was one of the four detectors operating at the LEP collider from 1989
to 2000.  It was designed as a general purpose detector for $e^+e^-$ physics
with special emphasis on precise tracking and vertex determination and on powerful particle 
identification.  A detailed
description of the DELPHI detector can be found in \cite{detector} and the
detector and trigger performance in \cite{performance,trigger}.

Charged particle tracks are reconstructed by a system of tracking chambers 
inside the 1.2~T solenoidal magnetic field: the Vertex Detector (VD), the Inner
Detector (ID), the Time Projection Chamber (TPC) and the Outer Detector
(OD) in the barrel region; two sets of planar drift chambers aligned
perpendicular to the beam axis (Forward Chambers A and B) measure tracks 
in the forward and backward directions.

The VD consists of three cylindrical layers of silicon detectors, at radii
6.3~cm, 9.0~cm and 11.0~cm, and polar angle acceptance from 24$^\circ$ to 156$^\circ$. 
All three layers measure coordinates in the plane
transverse to the beam ($xy$), and at least two of the layers also measure
$z$ coordinates along the beam direction.  
The ID consists of a cylindrical drift chamber with inner radius 12~cm and
outer radius 22~cm, surrounded by 5 layers of straw tubes, having 
a polar acceptance between 15$^\circ$ and 165$^\circ$.
The TPC, the principal tracking device of DELPHI, consists of a 2.7~m long cylinder of
30~cm inner radius and 122~cm outer radius. Each
end-plate of the TPC is divided into 6 sectors with 192 sense wires
and 16 circular pad rows per sector. The wires help in charged particle
identification by measuring the specific energy loss
(dE/dx) and the pad rows are used for 3 dimensional
space-point reconstruction. 
The OD consists of 5 layers of drift cells at radii between 192~cm and
208~cm, covering polar angles between 43$^\circ$ and 137$^\circ$.

The electromagnetic calorimeters consist of a High Density Projection Chamber
(HPC) covering the polar angle region from 40$^\circ$ to 140$^\circ$ and, a
Forward ElectroMagnetic Calorimeter (FEMC) covering the polar angle regions from 
11$^\circ$ to 36$^\circ$ and from 144$^\circ$ to 169$^\circ$.  The Scintillator
TIle Calorimeter (STIC) extends the polar angle coverage down to 1.66$^\circ$
from the beam axis in both directions. 
The Hadron CALorimeter (HCAL) covers 98\% of the solid
angle. The muons which traverse the HCAL are recorded in a set of Muon Drift
Chambers placed in the barrel, forward and backward regions.

The Ring Imaging CHerenkov (RICH) detectors of DELPHI~provide charged particle 
identification in both the barrel (BRICH)
and forward (FRICH) regions. In addition, a set of scintillation counters 
were added to veto photons in blind regions of the electromagnetic 
calorimeter at polar angles near 40, 90 and 140 degrees.

\section{Particle identification}
An isolated particle was identified as a muon if it gave signal in the muon
chambers or left a signal in the calorimeters compatible with a MIP\@.
It was identified as an electron if its energy deposition in the
electromagnetic calorimeters was compatible with its measured momentum and the
ionisation loss in the TPC was compatible with that expected from an electron
of that momentum.

If an electron or muon had low momentum, it was assumed to come from
$\tau$ decay and was therefore tagged as $\tau$. 
In addition isolated jets with an energy of at least 5 GeV, at least 
one and at most five charged tracks and no more than ten
particles in total were also considered as $\tau$ candidates. 

The $\tau$ decays were classified into the following categories:
$e$, $\mu$, $\pi$, $\pi+n\gamma$, $3\pi$ and others according to the lepton
identification, the number of charged tracks of the jet and the number of
photons.

\section{Topics}

Several searches with $\tau$ leptons in the final state are reviewed. Here, 
we will explain briefly the $\tau$ channels and will present the results 
with the combination of the rest of the possible decay channels.

\subsection{SM Higgs and MSSM neutral Higgs bosons}
The Standard Model (SM) Higgs search has been one of the most important goals at 
LEP2. 
A scalar Higgs could be produced at LEP in the process $e^+e^- \to HZ$. In the Minimal 
Supersymmetric Standard Model (MSSM), the production of the lightest scalar Higgs boson, h, 
proceeds through the same processes as in the SM.
There is also a CP-odd pseudo-scalar, A, which would be produced mostly in the 
$e^+e^- \rightarrow {\mathrm h} {\mathrm A}$ process at LEP2~\cite{higgshunter}. 

The BR($H \to \tau^+ \tau^-$) $\sim$5\%, then the $\tau^+\tau^-q\overline{q}$ channels 
corresponds to $\sim$9\% of the total decays. In the case of the MSSM neutral Higgs bosons the 
$\tau^+\tau^-b\overline{b}$ channels corresponds to $\sim$ 16\%. The same analysis was applied in 
the four channels with $\tau$ leptons~\cite{paper_higgs}. 

\begin{figure}[htb]
\includegraphics[width=20pc,height=20pc]{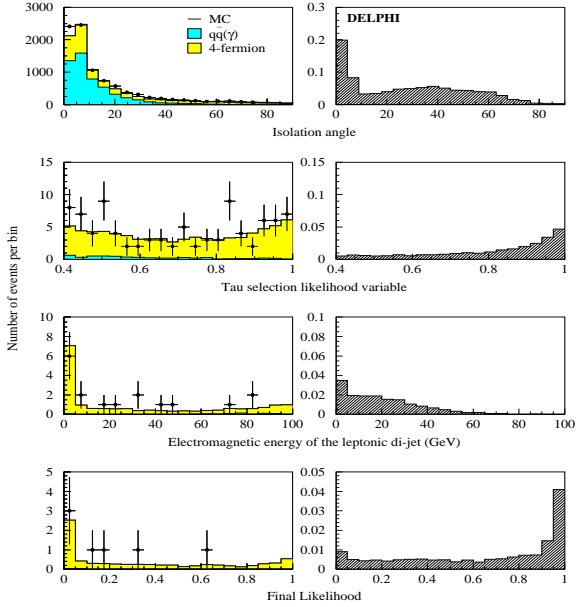}
\caption[]{Distributions of four analysis variables at different levels of the selection 
in the $\tau^+\tau^- q\overline{q}$ channel in the search for neutral Higgs bosons. 
Data from $\sqrt{s}$=~202-209~GeV (dots) are compared with SM background process
expectations (left-hand side histograms). The expected distribution for a 115~GeV/$c^2$ Higgs 
signal in the ($h Z \rightarrow \tau^+\tau^- q\overline{q}$) channel is shown in the 
right-hand side plots.}
\label{fig:htautau}
\end{figure}

After a hadronic preselection a search for $\tau$ lepton candidates was then performed using 
a likelihood ratio technique. The likelihood variable was calculated for the 
preselected particles using distributions of the particle momentum, of its isolation 
angle and of the probability that it came from the primary vertex.
Pairs of $\tau$ candidates were then selected requiring opposite charges, 
an opening angle greater than $90^\circ$ and a product of the
$\tau$ likelihood variables above 0.45. If more than one pair was
selected, only the pair with the highest product was kept. 
For the final selection another likelihood variable is built with the following variables: 
the rescaling factors of the $\tau$ jets, the $\tau$ momenta and the global 
\mbox{$b$-tagging} variable. Distributions of four analysis variables at different levels of 
the selection are shown in Fig.~\ref{fig:htautau}.

\begin{figure}[htb]
\includegraphics[width=18pc,height=15pc]{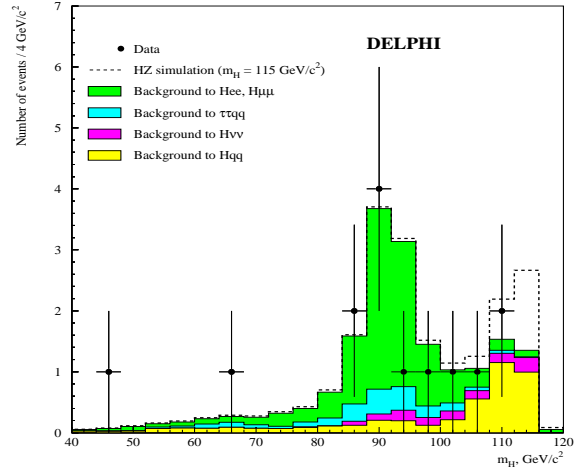}
\caption[]{Distribution of the sum of the reconstructed Higgs boson masses
for the tight candidates in each channel from the 2000 data in the SM Higss search.}
\label{fig:hzmass}
\end{figure}

For all searches analyses a likelihood ratio technique~\cite{alex} has been used to compute the cross-section and mass
limits. No evidence of Higgs boson production was found neither in the SM Higgs search nor the MSSM
neutral Higgs search. The distribution of the sum of the reconstructed Higgs boson masses
for the tight candidates in each channel from the 2000 data in the SM Higss search is shown in
Fig.~\ref{fig:hzmass}. With 
combination of all channels the lower limit at 95\% CL was set on the mass 
of the Standard Model Higgs boson at:

\begin{center}
$m_H > 114.1$ GeV/$c^2$
\end{center}

The following limits are derived in the framework of the MSSM:

\begin{center}
$m_h > 89.1$ GeV/$c^2$ and $m_A > 90.0$ GeV/$c^2$
\end{center}

\noindent
for all values of \tbeta\ above 0.44 and assuming 
the mixing in the stop sector.

\subsection{Charged Higgs bosons 2HDM type II}
The existence of a pair of charged Higgs bosons is predicted by several
extensions of the Standard Model. Pair-production of charged Higgs
bosons occurs mainly via $s$-channel exchange of a photon or a Z$^0$ boson. In two Higgs 
doublet models (2HDM),
the couplings are completely specified in terms of the electric charge and the weak mixing 
angle, $\theta_W$, and therefore the production cross-section depends only on the charged
Higgs boson mass. Higgs bosons couples to mass and therefore decay
preferentially to heavy particles, but the precise  branching ratios may vary
significantly depending on the model. In 2HDM type II, where the down/up fermions couple to 
$H_1$/$H_2$ Higgs fields respectively, the  $\tau \nu_\tau$ and $cs$ decay 
channels are expected to dominate. There are two channels with $\tau$ leptons in the
final state: the purely leptonic channel 
($H^+H^-\rightarrow \tau^+ \nu_\tau \tau^- \bar{\nu}_\tau$) and the semileptonic channels 
($H^+H^-\rightarrow q q' \tau \nu$)~\cite{paper_hphm}.

In both cases the main remaining background coming from $W$-pair production. Apart from the
reconstructed mass, which is not possible to reconstruct in the leptonic
channel, there are two
important difference to discriminate both contributions: the $\tau$ polarization and the boson 
production polar angle.

\begin{itemize}
\item $\tau$ polarization:

assuming that the $\nu_\tau$ has a definite helicity, the polarization
($P_\tau$) of tau leptons originating from heavy boson decays is
determined entirely by the properties of weak interactions and the
nature of the parent boson. The helicity configuration for the signal
is H$^- \rightarrow \tau^-_R \mbox{$\bar\nu_\tau$}^{\phantom +}_R$
(H$^+ \rightarrow \tau^+_L {\nu_\tau}^{\phantom +}_L$) and for the
W$^+$W$^-$ background it is W$^- \rightarrow \tau^-_L
\mbox{$\bar\nu_\tau$}^{\phantom +}_R$ (W$^+ \rightarrow \tau^+_R
{\nu_\tau}^{\phantom +}_L$) resulting in $P_\tau^{\rm H}=+1$ and
$P_\tau^{\rm W}=-1$. The $\tau$ weak decay induces a dependence
of the angular and momentum distributions on
polarization. Once the $\tau$ decay channel is identified, the
information on the $\tau$ polarization was extracted from the observed
kinematic distributions of its decay products, e.g. angles and
momenta.  These estimators are equivalent to those used at the Z peak for
precision measurements~\cite{ptauz}. For charged Higgs boson masses close to the
threshold, the boost of the bosons is relatively small and the $\tau$
energies are similar to the $\tau$'s from 
Z decays at rest (40--50~GeV).

\item Boson production polar angle:

$H$-pair production ocurrs via s-channel and the differential cross-section follows a behavior
proportional to $1+cos^2\theta$. But, $W$-pair production ocurrs via s-channel and t-channel. 
Then, the $W$ production polar angle distribution goes to $\cos \theta_W \sim 1$, and the 
$W^-$ ($W^+$) is emitted preferently in the direction of $e^-$ ($e^+$).
\end{itemize}

\begin{figure}[htb]
\includegraphics[width=20pc,height=17pc]{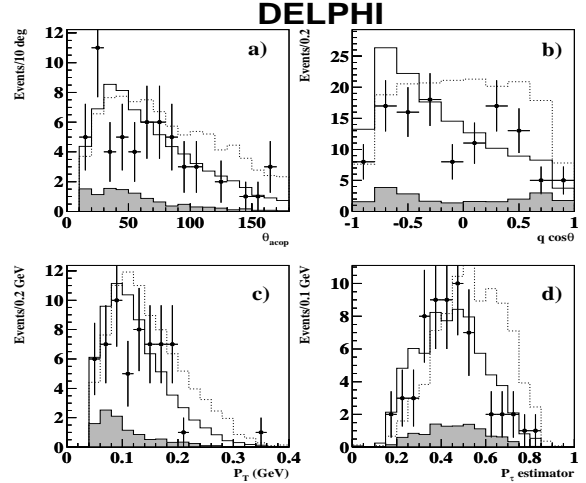}
\caption[]{
Distribution of the variables used for the anti-WW likelihood
 for the $\HTT$ analysis after preselection: 
  a) acoplanarity, b) cosine of polar angle accounting for the charge,
  c) transverse momentum escaled by the center-of-mass energy and d) $\tau$ polarization estimator.
  Data are shown as filled circles, while
  the solid histogram contour shows the expected SM 
  background with contributions from WW (unfilled) and other contributions (shaded).
  The expected histogram for a
  85 GeV/$c^2$ charged Higgs boson signal is shown as a dashed line
  in arbitrary normalization for comparison.}
\label{fig:fig_tntn_vars}
\end{figure}

In the leptonic channel, the selection is performed in similar way to the 
$W^+W^-\rightarrow \tau^+ \nu_\tau \tau^- \bar{\nu}_\tau$~\cite{delphiww}. After that 
selection, a likelihood to 
separate the signal from the background was built using six variables: the estimators of 
the $\tau$ polarization, the boson polar angle of both $\tau$ leptons, the acoplanarity 
and the total transverse momentum. These variables are shown in Fig.~\ref{fig:fig_tntn_vars}.

In the semileptonic channel two likelihood variables were built to reject the $QCD$ background
and the $WW$ background. In the anti-$WW$ likelihood the estimator of 
the $\tau$ polarization and the reconstructed polar angle of the negatively boson are included.

\begin{figure}[htb]
\includegraphics[width=20pc,height=15pc]{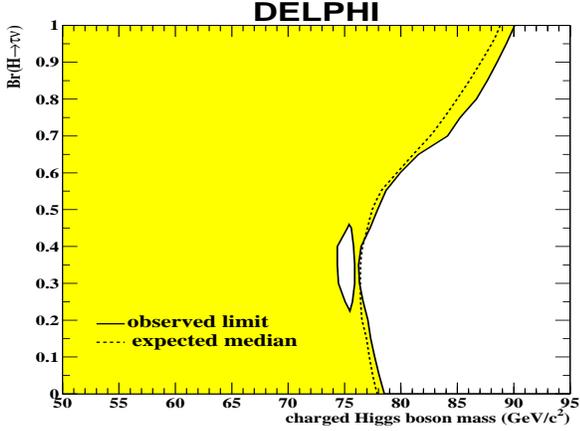}
\caption[]{
     The observed and expected exclusion regions at 95\% confidence level in the
     plane of BR($H \rightarrow \tau\nu_\tau$) vs. $M_{{\rm H}^{\pm}}$.
     These limits were obtained from a combination of the search results in the
      $\HTT$, $\HCSTN$ and $\HCSCS$ channels at  $\sqrt{s}=$ 183--209~GeV\@, under
     the assumption that the $\wa$ decay is forbidden.}
\label{fig:limithphm}
\end{figure}

No significant signal-like excess of events compared to the expected backgrounds 
was observed in any of the three final states investigated.  A lower limit for a charged Higgs 
boson mass was set at 95\% confidence level of $M_{{\rm H}^{\pm}} >$ 74.3~GeV/$c^2$, assuming 
BR($H \to \tau \nu$) + BR($H \to c s$) = 1. The observed and expected exclusion regions 
at 95\% confidence level in the plane of BR($H \rightarrow \tau\nu_\tau$) vs. 
$M_{{\rm H}^{\pm}}$ are shown in Fig.~\ref{fig:limithphm}. The noticeable 
difference between observed and expected limit is dominated by a small 
``hole'' around BR($H \to \tau \nu$)=0.35 which reaches only 92\% as 
confidence level, produced by a small excess of data in that region in 
the semileptonic channel.

\subsection{Charged Higgs bosons 2HDM type I}

In 2HDM type I models, where all fermions couple to the same Higgs doublet, the
\mbox{$H^- \to \wa$} decay can be dominant, if the neutral pseudoscalar $A$ is light and if $\tan\beta$ is large
enough~\cite{wateo}. To cover this eventuality the final states $\HWAWA$ and 
$\HWATN$ were also looked for~\cite{paper_hphm}. 

If at least one of the Higgs bosons decays to a $\wa$ pair, there are several possible
topologies depending on the different boson decays.  The W boson can decay leptonically or 
hadronically, and the number of jets strongly depends on the A mass and on the 
boson boosts. To treat all these decays in a generic way, the search was restricted to $A$
masses above 12 GeV, where it decays predominantly to $\bb$ and an inclusive
search is performed. Events with jets with $b$ quark content were
searched for in two topologies: events with a $\tau$, missing energy and at least two 
hadronic jets; and events with no missing energy and at least four hadronic jets. It was 
found that the analysis designed by DELPHI for technicolor
search\cite{delphitc} was well suited also for these topologies and had a good
performance on this search. It was therefore adopted here. In both topologies good 
agreement between the data and the Standard Model expectation was found.

\begin{figure}[htb]
\includegraphics[width=20pc,height=17pc]{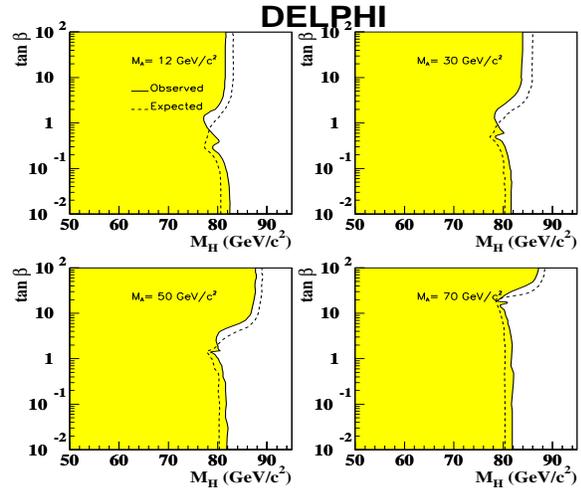}
\caption[]{
  The observed and expected exclusion regions at 95\% confidence level in the
  plane of  $\tan\beta$  vs. $M_{{\rm H}^{\pm}}$. These limits were
  obtained from a combination of the search results in all five 
  channels at $\sqrt{s}=$ 183--209~GeV\@, for different $A$ masses.} 
\label{fig:limitwa}
\end{figure}

Therefore, with the combination of the five channels, if the \mbox{$\wa$} decay is allowed, a 
lower H$^{\pm}$ mass limit of $M_{{\rm H}^{\pm}} >$ 76.7~GeV/$c^2$ can be set at the
95\% confidence level, independently of $\tan\beta$ for $M_A>12$. The observed and expected exclusion 
regions at 95\% confidence level in the plane of $\tan\beta$ 
vs. $M_{{\rm H}^{\pm}}$, for different $A$ masses, 
are shown in Fig.~\ref{fig:limitwa}.

\subsection{Technicolor search}
The technicolor model~\cite{tcteo} provides an alternative to the SM mechanism of Electroweak Symmetry breaking. 
Role of Higgs boson is performed by bound states of new fundamental fermions, Techniquarks, which mix
with $W$ and $Z$, and generate their masses. These bosons are seen as condensates of a new family of 
quarks (the techniquarks) which obey a QCD-like interaction with an effective scale 
$\Lambda_{TC}$ much larger than $\Lambda_{QCD}$. It also predicts heavy ($>$~1 TeV)
vector mesons which cannot be observed at LEP2. 

The main $\rho_T$ decay modes are $\rho_T \to \pi_T \pi_T$,
$W_L \pi_T$, $W_L W_L$, $f_i \bar{f_i}$ and $\pi_T^0 \gamma$,
where $W_L$ is the longitudinal component of the $W$ boson.
For $ M_{\rho_T} > 2 M_{\pi_T}$ the decay $\rho_T \to \pi_T \pi_T$
is dominant, while for $M_{\rho_T} < 2 M_{\pi_T}$ the decay rates depend on many model parameters. 
Technipions can also be produced at LEP through virtual $\rho_T$ exchange. 
Technipions are assumed to decay as $\pi_T^+ \to c \bar{b}$, 
$c \bar{s}$ and $\tau^+ \nu_{\tau}$; and $\pi_T^0 \to b \bar{b}$, 
$c \bar{c}$ and $\tau^+ \tau^-$. The width $\Gamma(\pi_T \to \bar{f'} f)$ 
is proportional to $(m_f +m_{f'})^2$, therefore the $b$-quark is 
produced in $\sim90$\% of $\pi_T$ decays. The total $\pi_T$ width is 
less than 1 GeV. The analyses performed by DELPHI use the off-shell processes
$e^+e^- \to \rho_T^* \to (\pi^+_T \pi^-_T$, $\pi^+_T W^-_L)$ and 
$e^+e^- \to (\rho_T^*, \omega_T^*) \to \pi^0_T \gamma$.

The search for the technipion~\cite{delphitc} in semileptonic channels containing 
two quarks, a lepton and a neutrino, corresponds to the decays 
$W_L^+ \pi_T^- \to l^+ \nu_l b \bar{c}$ and $\pi_T^+ \pi_T^- \to 
\tau^- \bar{\nu_\tau} b \bar{c}$. The first step in the analysis was a $qq'l\nu$ selection.
The second step exploits the specific properties of the signal, such as the presence of 
b-quarks or the production angle, to distinguish it from the W pairs.
This is done building a neural network which uses four input variables: 
the $b$-tagging variables of the two hadronic jets, 
$\cos{\theta_{W^-}}$ and $|\cos{\theta_{miss}}|$.

\begin{figure}[htb]
\includegraphics[width=20pc,height=17pc]{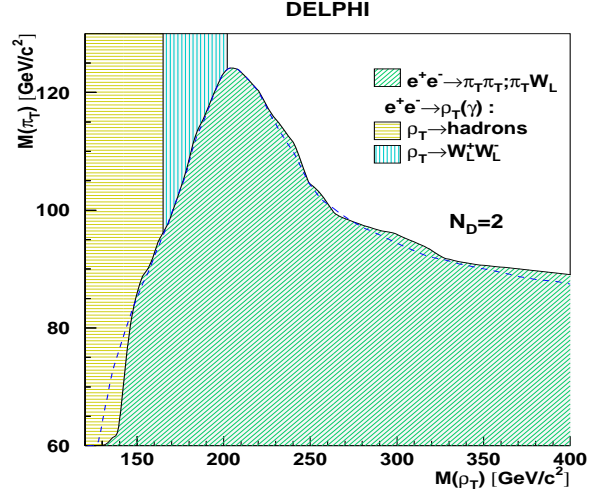}
\caption[]{The region in the $(M_{\rho_T} - M_{\pi_T})$ plane
      (filled area) excluded at 95\% CL for $N_D=2$ (maximal $W_L$ - $\pi_T$
      mixing). The dashed line shows the expected limit for the 
      $e^+e^- \to \pi_T \pi_T, \pi_T W_L$ search.}
\label{fig:limit_tcnd2}
\end{figure}

Good agreement between the data and the SM background was observed 
in all channels studied in this search. 
The combined region in the $(M_{\rho_T},M_{\pi_T})$ plane excluded 
by this analysis at a 95\% CL is shown for $N_D=2$ (maximal $W_L$ - $\pi_T$ mixing) in 
Figure~\ref{fig:limit_tcnd2}.

\subsection{Invisibly decaying Higgs bosons}
Some extensions of the SM exhibit Higgs bosons which can decay into stable neutral weakly
interacting particles, therefore giving rise to invisible final states. In supersymmetric
models the Higgs bosons can decay with a large branching ratio into lightest neutralinos or
gravitinos in some region of the parameter space, leading to an invisible final state if
$R$ parity~\footnote{R-parity is a multiplicative quantum number defined as 
$R_P=(-1)^{3B+L+2S}$ where B, L and S are the baryon number, the lepton number and the spin of
the particle, respectively. SM particles have $R=+1$, while their SUSY particles have $R=-1$.} 
is conserved~\cite{djouadi,peccei,campos,wells}.

Searches for {\mbox{$ {\mathrm H} {\mathrm Z}$}} production with the Higgs boson 
decaying into an invisible final state were performed by the DELPHI experiment. 
Both hadronic and leptonic final states of the Z boson were analyzed. DELPHI 
has been analyzed the four possible visible final states channels, 
including the $\tau^+\tau^-$ channel.

The selection is explained in detail in~\cite{paper_invhiggs}, and it was performed in similar
way as other searches with two $\tau$ leptons plus additionals non-interacting particles. 
The mass of the invisibly decaying particle was computed from the measured
energies assuming momentum and energy conservation. In the case of the $\tau^+\tau^-$ channel, the information carried by
the decay products does not reproduce correctly the $\tau$ energy. 
Therefore, the mass is calculated under the assumption that both $\tau$ 
leptons had the same energy.

\begin{figure}[htb]
\includegraphics[width=18pc,height=15pc]{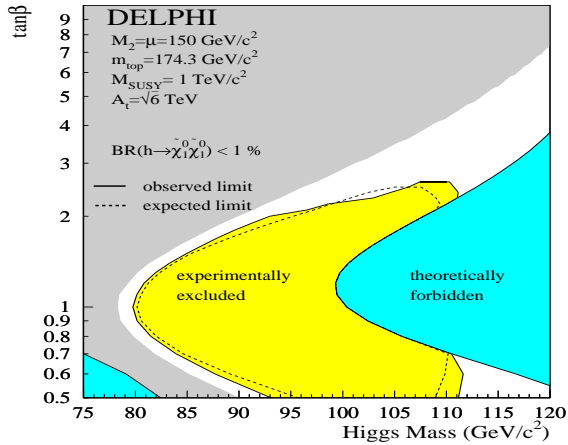}
\caption[]{Excluded region in the MSSM from searches for 
Higgs boson decaying into invisible final states
for $\mathrm{m_h}$-max in the stop sector.
The different grey areas show the theoretically forbidden region (dark),
the region where the Higgs boson does not decay into neutralinos 
(intermediate),
the region which is excluded at 95\% CL by this search for invisibly 
decaying Higgs bosons (light) and the unexcluded region (white).}
\label{fig:mssmmax}
\end{figure}

From the comparison with the SM Higgs boson cross-section, and combining the four final states
investigated, the observed (expected) mass limits are 112.1 (110.5)~GeV/$c^2$
for the Higgs boson decaying into invisible particles. The excluded regions in the MSSM from 
searches for  Higgs boson decaying into invisible final states
for $\mathrm{m_h}$-max in the stop sector are shown in Fig.~\ref{fig:mssmmax}.

\subsection{SUSY searches}
Supersymmetry (SUSY) is at present one of the most attractive possible extensions of the SM and
its signatures could be observed at LEP through a large variety 
of different channels~\cite{susyteo}.
Supersymmetry (SUSY) is usually assumed to be broken in a hidden sector of particles
and then communicated to the observable sector (where all the particles and
their superpartners lie) via gravitational interactions. But there are other possibilities 
where this mediation is performed by SM gauge interactions, leading to models of 
gauge mediated supersymmetry breaking.

Many channels and several scenarios have been studied with $\tau$ leptons in the final state. 
DELPHI have performed searches for supersymmetric particles within the framework of the 
MSSM assuming R-parity conservation~\cite{paper_susyrpc} 
(implying that the Lightest Supersymmetric Particle, LSP, 
is stable and SUSY particles are pair-produced), with R-parity 
violation~\cite{paper_susyrpv}, in Gauge Mediated Supersymmetry Breaking (GMSB) 
models~\cite{paper_susygravitino} and in Anomaly Mediated SUSY Breaking (AMSB) 
models~\cite{paper_amsb}.

As an example, the different topologies studied in the GMSB framework are summarized in 
Table~\ref{tab:gmsb}. The $\tau$ leptons appear in several channels: acoplanar 
leptons (2 leptons $+$ missing energy), kinks and large impact parameters (when the 
sparticle decays inside the detector) and four leptons $+$ missing energy (if 
$\tilde\tau$ is the Next Lightest Supersymmetric Particle, then there are 
4$\tau$ in the final state).

\begin{table}[h!]
  \begin{center}
    \begin{tabular}{|l|l|l|}
      \hline
      Production & Decay mode & $\hat{L}$  \\
      \hline
      \hline
      &  & $ << l_{detector}$   \\
      $ \sLep \sLep$ & $\sLep \to l \sGra$   &  $\sim l_{detector}$  \\
      &  &  $ >> l_{detector}$    \\
      \hline
      $ \Chiz_1 \Chiz_1$ & $\Chiz_1 \to \sLep l \to l l \sGra$  &  $<< l_{detector}$  \\
      \hline
      & &  $ << l_{detector}$    \\
      $\Chip_1 \Chim_1$ & $\Chip_1 \to \sLep^+
      \nu \to l^+ \sGra \nu$  &  $\sim l_{detector}$ \\
      &  & $ >> l_{detector}$   \\
      \hline
      $\phi \gamma$ & $\phi \rightarrow \gamma\gamma$ & $ <<
      l_{detector}$  \\
      & $\phi \rightarrow gg$ & $ << l_{detector}$ 
       \\ \hline
    \end{tabular}
  \end{center}
\caption[]{Final state topologies studied in the different scenarios, within the GMSB models.}
\label{tab:gmsb}
\end{table}

\begin{figure}[htb]
\includegraphics[width=20pc,height=17pc]{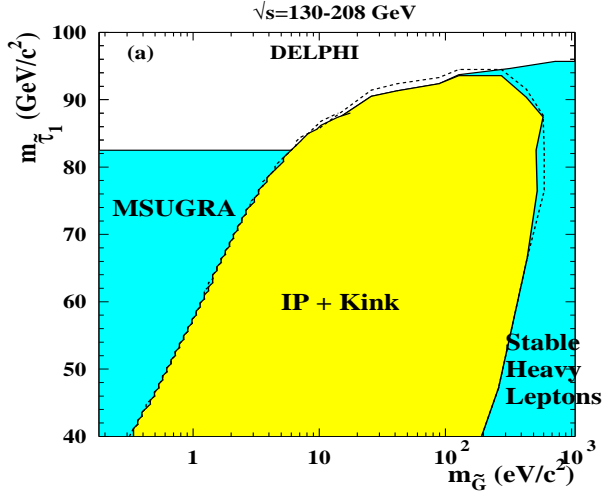}
\caption[]{Exclusion regions in the 
($m_{\tilde{G}}$,$m_{\tilde{\tau}_1}$) plane at 
95\%~CL combining all analyses in GMSB models.}
\label{fig:exlcgmsb}
\end{figure}

In all cases no evidence for signal production was found. Hence, the DELPHI collaboration set
lower limit at 95\% C.L. for the sparticles masses. Exclusion regions in the 
($m_{\tilde{G}}$,$m_{\tilde{\tau}_1}$) plane at 
95\%~CL combining all analyses in GMSB models are shown in Fig.~\ref{fig:exlcgmsb}.

\subsection{Doubly charged Higgs bosons}
Doubly charged Higgs bosons ($H^{\pm\pm}$) appear in several extensions 
to the Standard Model~\cite{hpphmmtheory}, such as
left-right symmetric models, and can be relatively light. 
In Supersymmetric left-right models usually the $SU(2)_R$ gauge symmetry is
broken by two triplet Higgs fields, so-called left and right handed. 
Pair-production of doubly charged 
Higgs bosons is expected to occur mainly via $s$-channel 
exchange of a photon or a \rm{Z} boson. In left-right symmetric models 
the cross-section of $e^+e^- \to H^{++}_L H^{--}_L$ is different from that for 
$e^+e^- \to H^{++}_R H^{--}_R$, where $H^{\pm\pm}_L$ and $H^{\pm\pm}_R$ are the
left-handed and right-handed Higgs bosons. 

The dominant decay mode of 
the doubly charged Higgs boson is expected to be 
a same sign charged lepton pair, the decay proceeding via a lepton number 
violating coupling. As discussed in~\cite{hpphmmlimits}, due to limits that 
exist for the couplings of 
$H^{\pm\pm} \to e^\pm e^\pm$ from high energy Bhabha scattering, 
$H^{\pm\pm} \to \mu^\pm \mu^\pm$ from the absence of muonium to anti-muonium
transitions and $H^{\pm\pm} \to \mu^\pm e^\pm$ from limits on the flavour
changing decay $\mu^\pm \to e^\mp e^\pm e^\pm$, electron and muon decays 
are not likely. In addition, most of the models expect that the coupling to
$\tau\tau$ will be much larger than any of the others. Therefore, 
only the doubly charged Higgs boson decay $H^{\pm\pm} \to \tau^\pm \tau^\pm$ 
is considered by DELPHI~\cite{paper_hpphmm}.

Depending on the $h_{\tau\tau}$ coupling and the Higgs
mass the experimental signature is different. If $h_{\tau\tau}$ is 
sufficiently large, $h_{\tau\tau} \geq 10^{-7}$, 
the Higgs decays very close to the interaction point. If $h_{\tau\tau}$ is smaller 
the decay occurs inside the tracking detectors (the analysis used in the search for kinks in 
GSMB models is adopted here) or even beyond them (search for stable 
massive particles). In the first case, the resulting final state 
consists of four narrow and low multiplicity $\tau$ jets coming from the interaction 
point. The analysis was performed in several steps: first a four jet preselection was applied,
followed by some cuts to reject the 2-photon and the 4-lepton background. Finally the mass was 
built requiring energy and momentum conservation. After all requirements were applied only one 
event was observed in the data, while 0.9 events were expected from background processes.

\begin{figure}[htb]
\includegraphics[width=20pc,height=17pc]{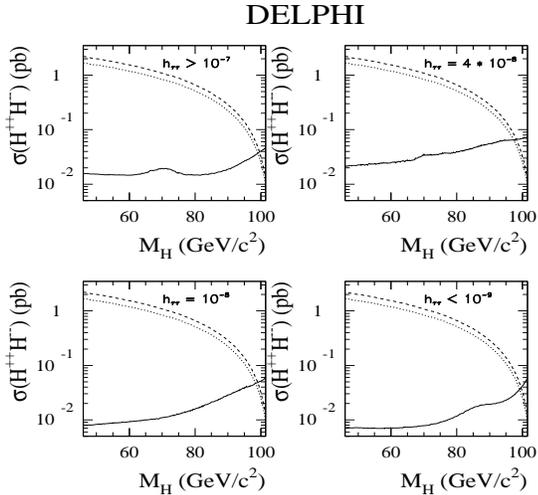}
\caption[]{The solid line shows the 95\% confidence level upper limit on the
$H^{\pm\pm}$ pair production cross-section at $\sqrt{s}$=206.7 GeV assuming
100\% branching ratio for the decay of $H^{\pm \pm}$ into $\tau^\pm\tau^\pm$ for
different values of $h_{\tau\tau}$. 
The dashed and dotted lines show the expected production cross-section of 
$H^{\pm \pm}_L$ and $H^{\pm \pm}_R$ pairs in left-right symmetric models.}
\label{fig:xslimit}
\end{figure}

The three different analyses were applied to cover the whole range of the $h_{\tau\tau}$ 
coupling, and good agreement between the data and the SM background was observed. 
Fig.~\ref{fig:xslimit} shows the 95\% confidence level upper limits on 
the cross-section at $\sqrt{s}$ = 206.7 GeV for
the production of $H^{++}H^{--} \to \tau^+\tau^+\tau^-\tau^-$ for different values of 
$h_{\tau\tau}$. The comparison of these limits with the expected cross-section for left-handed 
$H^{\pm \pm}_L$ and right-handed $H^{\pm \pm}_R$ pair production yields 
95\% confidence level lower limits for any value of the $h_{\tau\tau}$ coupling 
on the mass of the $H^{\pm\pm}_L$ and $H^{\pm\pm}_R$ bosons of 98.1 
and 97.3 GeV/$c^2$ respectively.

\subsection{Searches for Neutral Higgs Bosons in Extended Models}
The simplest extensions of the Standard Model are
the so-called Two-Higgs Doublet Models (2HDM), of which various types
exist, depending on the choice of the scalar couplings to
fermions~\cite{higgshunter}. The first type assumes that one doublet only couples to
fermions while the other one couples to gauge bosons. The resulting
final states display Higgs boson decays into photon pairs. 
The second and most studied type assumes
that one doublet couples to the up-type fermions (neutrinos and the
u,c,t quarks) while the other one couples to down-type fermions
(charged leptons and the d,s and b quarks). Depending on the mixing of
the two doublets, the dominant Higgs boson decays will be either
c-quarks and/or gluons, or b-quarks and $\tau$-leptons.

Finally a third possible choice of couplings, in which
one Higgs doublet couples to leptons only, while the other couples to
quarks. In this unexplored case, the dominant Higgs boson decay modes
may be leptonic, leading, in the case of Higgs boson pair production,
to the striking 4-$\tau$ final state. For the first time at LEP2 a search 
for $h A \to \tau^+ \tau^- \tau^+ \tau^-$ was performed at DELPHI~\cite{paper_haextended}. 
When the Higgs mass was lower than $\sim$25 GeV/$c^2$ the 
$\tau$ pairs of the Higgs decay were observed as only one jet due to the 
low angle between both $\tau$ leptons. Then, three different analyses 
were applied to keep the efficiency in a wide range of masses: '4-jet
analysis', '3-jet analysis' and '2-jet analysis'.

Good agreement between the data and the expected background was observed for the 
combined sample, and the combination of the three independent analyses. At the final 
selection level, 13 events were selected in the data and 17.9 events were expected 
from the Standard Model background. Very strong constrains on the 
$\tau$ coupling have been set with this analysis.

\section{Conclusions}
$\tau$ leptons are a sensitive probe for Physics beyond the Standard Model. A lot of 
new particle searches have been performed in the DELPHI experiment involving $\tau$ 
leptons in the resulting final state. New techniques have been used to discriminate 
the signal and the background.

Good agreement between the data and the Standard Model prediction has been 
found in all of cases. Therefore, limits on new particles have been set, playing 
an important role the channels with $\tau$ leptons.

\end{document}